\documentclass[aps,prl,twocolumn,showpacs]{revtex4}
\usepackage{amssymb}
\usepackage{amsmath}
\usepackage[dvips,pdftex]{graphicx}
\usepackage{color}

\begin{document}

\title{Correlated metallic two particle bound states in quasiperiodic chains}

\author{Sergej Flach${}^{1}$}
\author{Mikhail Ivanchenko${}^{1,2}$}
\author{Ramaz Khomeriki${}^{1,3}$}
\affiliation{ ${\ }^1$Max-Planck Institute for the Physics of
Complex Systems,
N\"othnitzer Str. 38, 01187 Dresden, Germany \\
${\ }^2$Theory of Oscillations Department, University of Nizhniy
Novgorod,
Russia \\
${\ }^3$Physics Department, Tbilisi State University,
Chavchavadze 3, 0128 Tbilisi, Georgia}

\begin{abstract}
Single particle states in a chain with quasiperiodic potential show a metal-insulator transition
upon the change of the potential strength. We consider two particles with local interaction
in the single particle insulating regime. The two particle states change from being localized to
delocalized upon an increase of the interaction strength to a nonperturbative finite value.
At even larger interaction strength
the states become localized again. This transition of two particle bound
states into a correlated metal is due to a resonant mixing of the noninteracting two particle eigenstates.
In the discovered correlated metal states two particles move coherently together
through the whole chain, therefore contributing to a finite conductivity.
\end{abstract}

 \maketitle

The interplay of localization and many body interactions has been a highly
active research topic since the discovery of Anderson localization \cite{PWA58}.
While the direct theoretical study of such systems is quite complex and recently hotly
debated (see e.g. \cite{dmbilabla06}), another route is taken by studying few interacting
particles. The potential applicability to recent experimental activities
with ultracold atoms in optical lattices \cite{immbloch} increased the interest in
corresponding theoretical studies. Localization of single particle states in a one-dimensional
lattice can be achieved using a lattice potential which is
biased by
an external dc field leading to Wannier-Stark localization and Bloch oscillations \cite{BO},
random leading to Anderson
localization \cite{PWA58}, and quasiperiodic leading to Aubry-Andre localization
\cite{aubry}.
Two interacting particles (TIP) in a dc field lead to no substantial change of the localization
length due to the Stark ladder structure of the single particle eigenenergies (disregarding some
resonant tunneling events which however are also suppressed for large distances) \cite{kkf09}.
TIP in the random case do lead to an increase of the localization length
with some controversial discussions about the quantitative outcome \cite{Shep_94,Imry_95,kkf11},
but do not yield complete delocalization,
For TIP in the quasiperiodic potential case few numerical results give varying predictions
from incremental
increase of localization length to opposite reports of decrease of localization length in the
insulating regime
\cite{Shep_96,Shreib,Evan}.

In this work we consider the TIP problem in a quasiperiodic chain at finite (nonperturbative)
strength of interaction, deep in the single particle insulating regime. We observe a complete delocalization
of certain two particle bound states, in which both particles are keeping a relative distance less
than the one particle localization length. The interaction renormalizes eigenenergies of different
classes of localized
TIP eigenstates with different strength, and leads to a resonant overlap of these energies
in a certain range of the interaction constant. In this nonperturbative window the overlap
between these groups of eigenstates leads finally to a complete delocalization. This results in
the novel observation of a correlated metal state build from only two interacting particles.

We study the TIP in the framework of the Hubbard model with Hamiltonian
\begin{equation}
{\cal \hat H}= \sum\limits_j\left[\hat b_{j+1}^+\hat b_j+\hat
b_{j}^+\hat b_{j+1}+\epsilon_j\hat b_{j}^+\hat b_j+\frac{U}{2}\hat
b_{j}^+\hat b_j^+\hat b_{j}\hat b_j\right] \label{eq1}
\end{equation}
where $\hat b_{j}^+$ and $\hat b_{j}$ are creation and
annihilation operators of indistinguishable bosons at lattice site
$j$, and $U$ measures the onsite interaction strength between the bosons.
The potential
\begin{equation}
\epsilon_j=\lambda\cos(\beta+2\pi\alpha n)  \label{eq11}
\end{equation}
controls the single particle problem via its strength $\lambda$.
The incommensurability parameter $\alpha=(\sqrt{5}-1)/2$ (golden mean).
$\beta$ is an arbitrary phase which controls wave packet dynamics and is
irrelevant for the properties of extended metallic eigenstates.

For a single particle the interaction term does not contribute.
Using the basis $|j\rangle\equiv\hat b_j^+|0\rangle$ the
eigenstates $|\nu\rangle$ of Hamiltonian
\eqref{eq1} with eigenvalues $\lambda_\nu$ are computed using
$|\nu\rangle=\sum\limits_j A_j^\nu b_j^+|0\rangle\;,\; {\cal \hat
H}|\nu\rangle=\lambda_\nu |\nu\rangle$.
$A_j^\nu\equiv \langle j|\nu\rangle$ are the eigenvectors.
As was first shown by Aubry and Andre \cite{aubry} (see also \cite{fishman,hang}),
for $\lambda<2$ all
eigenstates are delocalized (metallic phase) and for $\lambda>2$
all states are localized (insulating phase).
with a
localization length $\xi_1=1/\ln(\lambda/2)$. Consequently the probability
distribution function (PDF) $p_l^{(\nu)}=\langle q | \hat b_{l}^+\hat b_l |
q\rangle\sim\exp{[-2|l|/\xi_1]}$ is exponentially localized in
the insulating regime.  The bounded eigenvalue spectrum shows
fractal properties (see e.g. \cite{shep3}. At the value $\lambda=2.5$ (which is the main reference
parameter for computational studies in this work)
the spectrum has three main
minibands called SP1, SP2, and SP3 with centers around $\lambda=-2.5,0,2.5$ and ordered with increasing energy.
\begin{figure}[b]
\includegraphics[angle=90,width=1\columnwidth]{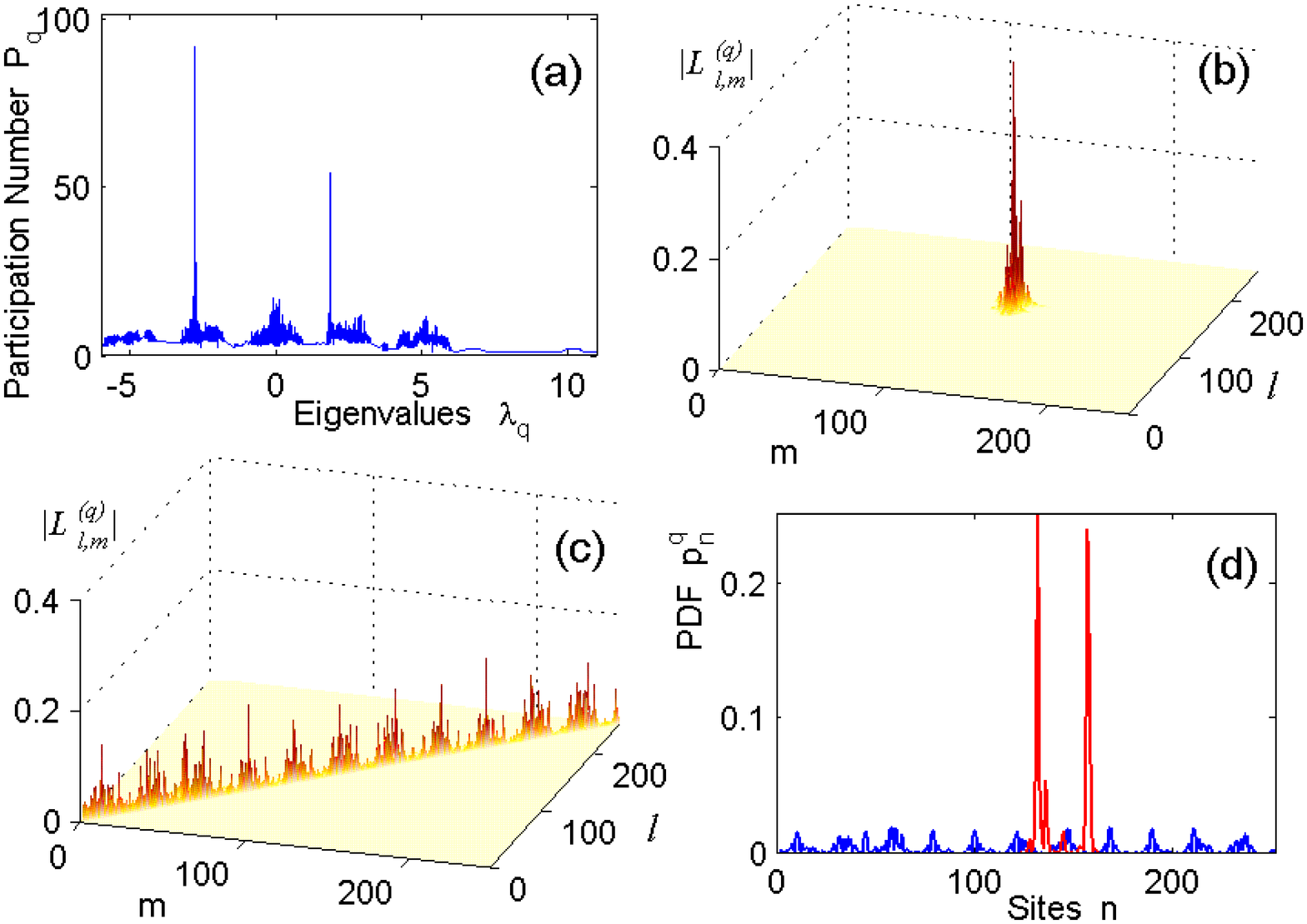}
\caption{(a) Participation number of interacting particle
eigenstates versus their eigenenergies, (b) and (c) present
typical localized and delocalized eigenfunctions $|{\cal
L}_{l,k}^{(q)}|$ which correspond to the background and peaks
in graph (a), respectively. In (d) we plot PDFs of
localized (red) and delocalized (blue) eigenmodes. Here
$U=7.9$ and $\lambda=2.5$.} \label{fig_4}
\end{figure}

For two particles we expand the eigenstates $| q\rangle$ of the
TIP problem in the local basis
\begin{equation}
|q\rangle=\sum_{m, l\le m}^N{\cal L}_{l,m}^{(q)}| l,m \rangle,
\quad  |l,m \rangle\equiv
\frac{b_{l}^+b_{m}^+|0\rangle}{\sqrt{1+\delta_{lm}}},
\label{eq_plh}
\end{equation}
where ${\cal L}_{l,m}^{(q)}=\langle l,m | q\rangle$ are the
normalized eigenvectors with $l\leq m$. Thus $[{\cal
L}_{l,m}^{(q)}]^2$ is the probability to find two
particles on the sites $l$ and $m$. We also
compute the PDF of the
eigenmodes $|q\rangle$ as
\begin{equation}
p_l^{(q)}=\frac{\langle q | \hat b_{l}^+\hat b_l |
q\rangle}{2}=\dfrac{1}{2}\left(\sum_{k, l\le k}^N{\cal
L}_{l,k}^{(q)2}+ \sum_{m, l\ge m}^N {\cal L}_{m,l}^{(q)2}\right)
\label{eq_pl}
\end{equation}
and  their participation numbers
$P_q=1\bigr/\sum_l^N\left(p_l^{(q)}\right)^{2}$.
We note that at the noninteracting limit $U=0$ and $\lambda=2.5$ the spectrum decomposes into
five minibands TP1,TP2,TP3,TP4,TP5 (see Fig. \ref{fig_4} below). This follows from
the three miniband structure of the single particle spectrum, such that e.g.
TP1 is formed from two single particle states SP1: $TP1=(SP1\times SP1$). Analogously,
$TP2=(SP1\times SP2), \ TP3=(SP2\times SP2)\cup(SP1\times SP3), \ TP4=(SP2\times SP3), \ TP5=(SP3\times SP3)$,
where $TPp=(SPm\times SPn)$ corresponds to a two-particle product state TPp with one of the particles being
in a single-particle eigenstate in miniband SPm and
another one in SPn.

To check for the selective character of a delocalization effect
we present participation numbers of all
eigenstates versus their eigenenergies for the particular choice of
interaction constant $U=7.9$ (Fig. \ref{fig_4}(a)).
The five miniband structure of the noninteracting case is clearly seen.
This follows from the fact, that most states correspond to two particles separated
by a distance larger than the localization length, and therefore these states
do not change when the interaction is increased. It is these states which also stay at a small
value of the participation number of the order of $P \approx 5$. However in the minibands TP2 and TP4 we observe
candidates for metallic delocalized states with $50 < P < 100 $ being one order of magnitude larger. Two characteristic eigenvectors
for localized and delocalized states are shown in Fig.\ref{fig_4}(b,c), and their corresponding
PDFs are plotted in Fig.\ref{fig_4}(d). The metallic state in Fig.\ref{fig_4}(b,d) is indeed
occupying the whole system. For this state both particles stay close to each other, forming
a diagonal structure in Fig.\ref{fig_4}(b). Therefore we coin these correlated
states metallic two particle bound states.

\begin{figure}[t]
\includegraphics[angle=0,width=1\columnwidth]{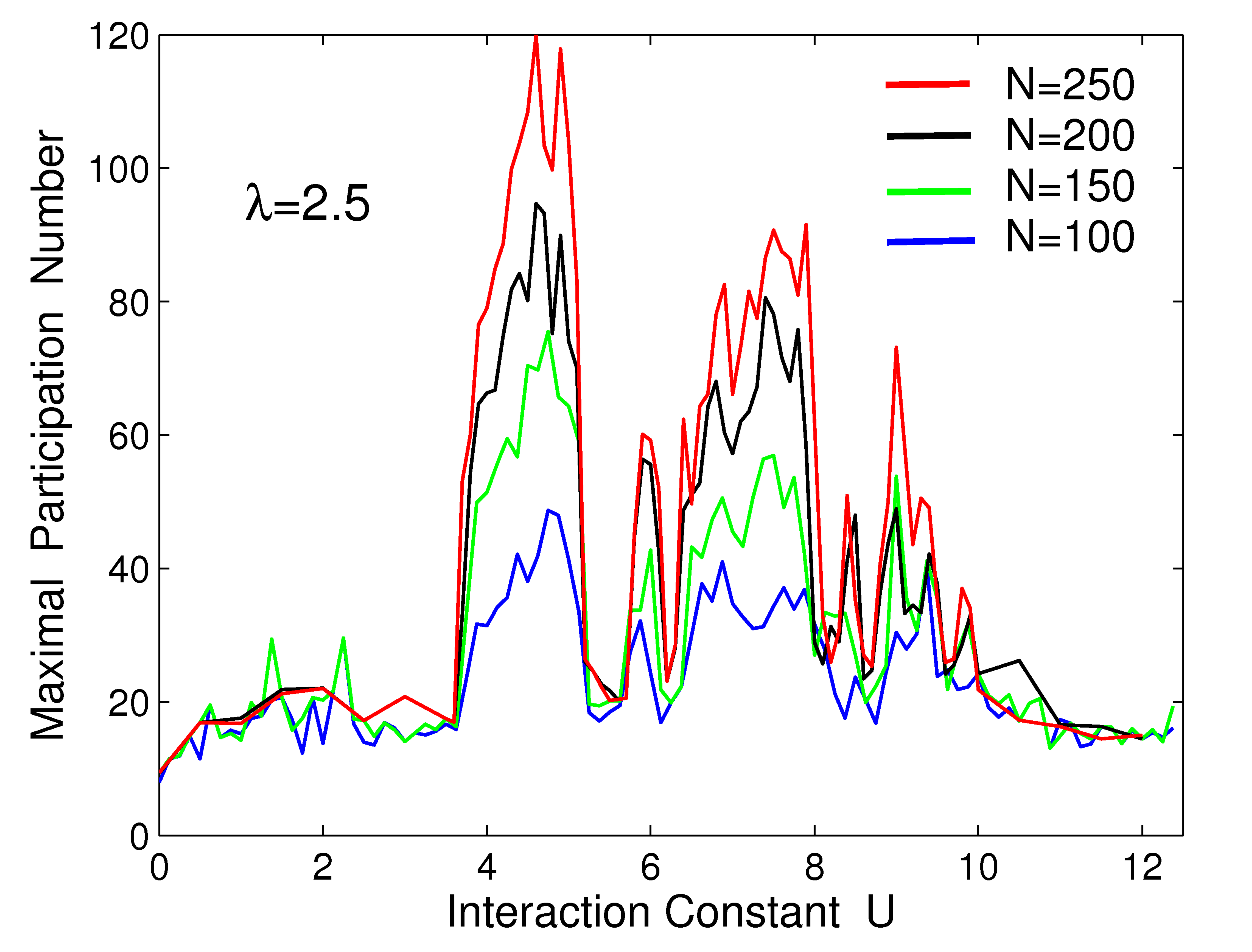}
\caption{Dependence of the largest participation number (of the mostly delocalized
eigenstate) on $U$ for different lattice sizes $N=100,150,200,250$ (from
bottom to top). Here $\lambda=2.5$.} \label{fig_3}
\end{figure}

Next we examine the
characteristics of eigenstates for $\lambda=2.5$ and various systems sizes and different values
of $U$.
We start with $N=100$. For each value of $U$ we find the state with the largest
participation number and plot this number versus $U$ in Fig.\ref{fig_3}.
While for $U < 4$ and $U> 10$ the number stays around 10-20, three humps up to values of 40 are noted
for $4 < U < 10$. Since the spatial size of an eigenvector is roughly 2-3 times larger than its
participation number, we conclude that in the hump regions the analyzed states extend over the
whole system. We increase the system size to $N=150,200,250$ and repeat the above analysis.
We observe that the hump heights grow linearly with the system size, indicating that for
all system sizes the most delocalized eigenstates in the hump regions are spreading over the whole
system. This is a clear evidence of complete delocalization of some eigenvectors in
the mentioned parameter region.

\begin{figure}[t]
\includegraphics[angle=0,width=1\columnwidth]{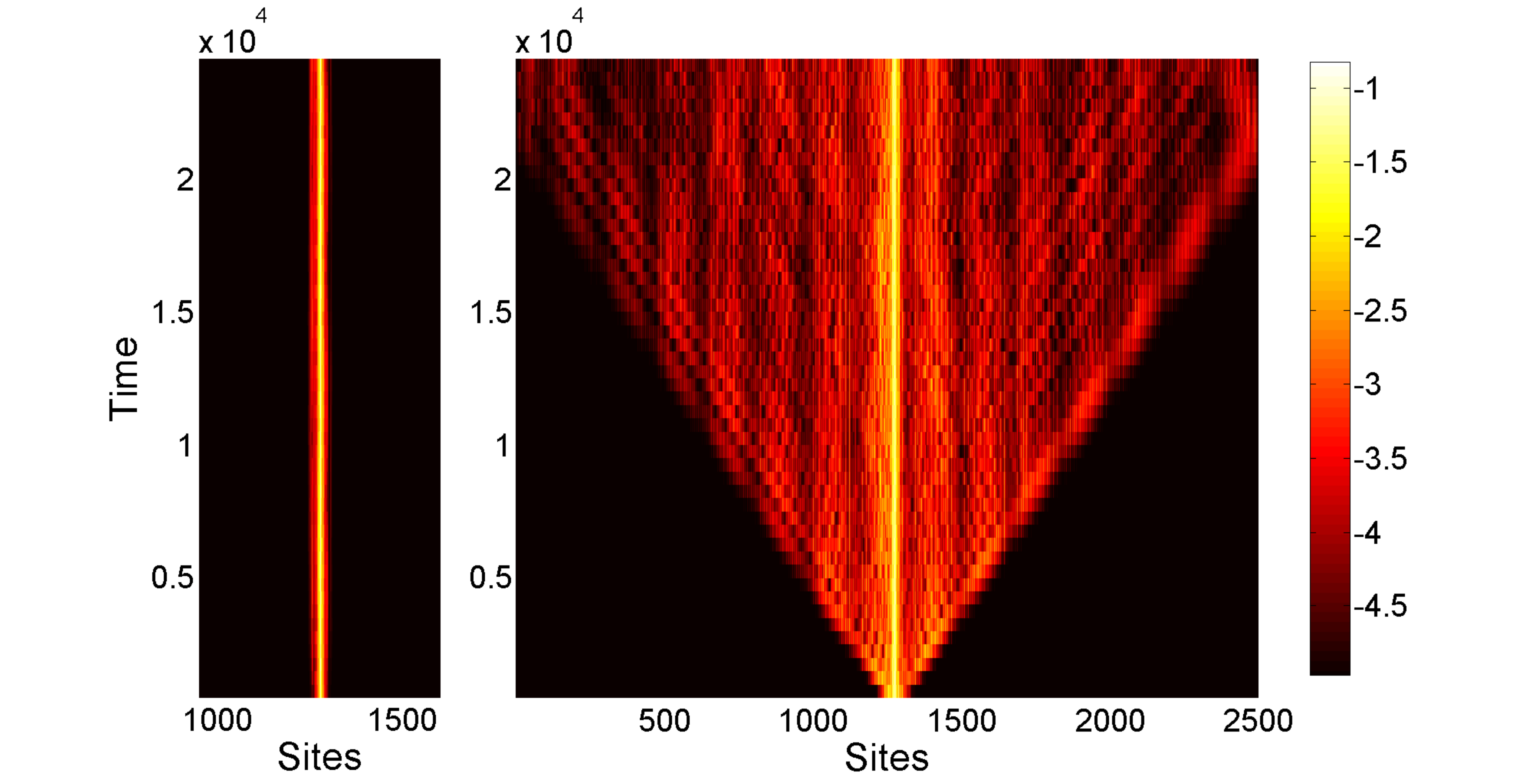}
\caption{(Color online) Time and space dependence of the PDF of an initial state with
two particles at adjacent sites for $\lambda=2.5$ (color maps $\log_{10}PDF$).  Left panel: isolator phase, $U=2$,
right panel: correlated metal phase, $U=4.5$.} \label{fig_2}
\end{figure}

Using exact diagonalization we are restricted to lattice sizes $N < 300$.
In order to push the limits, we compute the evolution of the Schr\"odinger
equation in real time without diagonalization, starting with two particles
located on adjacent sites. When and if the extended states exist, such initial conditions must excite them,
making a part of a two-particle wave packet propagate. Typical PDFs as a function of space and time are shown
in Fig.\ref{fig_2}.
We find ballistic spreading over the whole system with $N=2500$ sites for $U=4.5$,
and complete localization for $U=2$. This extends the evidence for complete delocalization
of the initial state into extended eigenstates of two interacting particles.

\begin{figure}[t]
\includegraphics[angle=0,width=1\columnwidth]{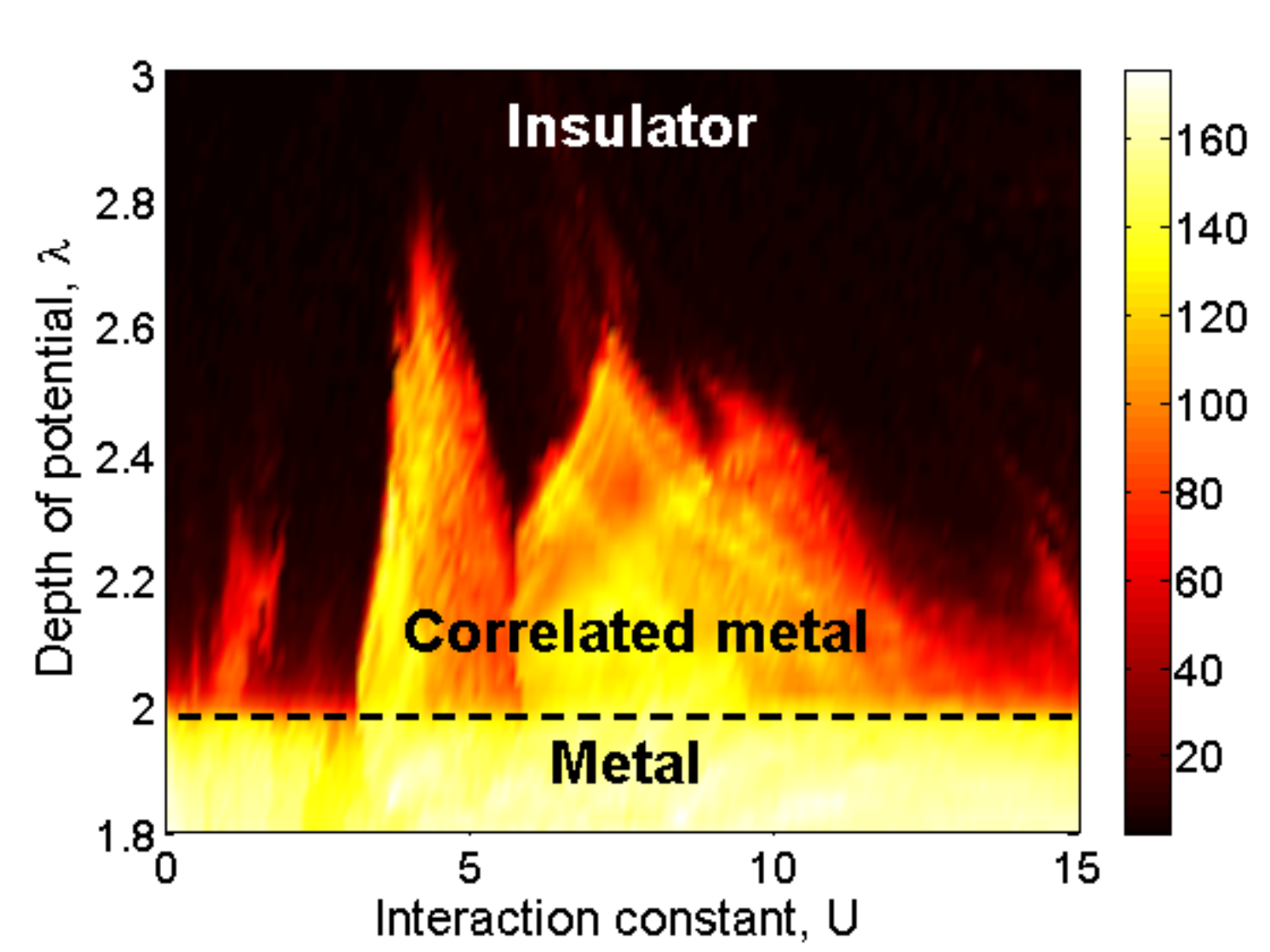}
\caption{ (Color online) The phase diagram of the TIP in a
quasiperiodic potential as suggested by the long term wave packet
evolution (the color code maps its second moment square root, see
text for further details). In light areas metallic eigenstates
exist, in dark -- localized eigenstates only. Dashed horizontal
line marks the MIT for the noninteracting case. Below it the
system is a metal. For nonzero interactions a new phase of a
correlated metal is formed in the midst of the insulator region. }
\label{fig_4a}
\end{figure}

We perform a scan in the parameter space $\{U,\lambda\}$ in order
to identify the region of correlated metallic two particle bound
states. We choose a system size $N=610$, and place two particles
at adjacent sites in the center of the chain. The Schr\"odinger
equation is evolved up to time $t=1.5\times 10^4$. The square root
second moment of the wave packet PDF is then measured for $60$
different original particle positions and the outcome for the
fastest growing realization is plotted in color code as a function
of both $\lambda$ and $U$ in Fig.\ref{fig_4a}. For $\lambda \leq
2$ all single particle states are metallic, therefore this region
is not of interest. However for $\lambda > 2$ single particle
states are localized. Here we find a large region of metallic
bound states for $3 \lesssim U \lesssim 15$ and $\lambda \lesssim
3$. Towards larger values of $\lambda$ the existence region breaks
up into two main tongues, which we observed also in
Fig.\ref{fig_3}. Note also two separate tongue structures at
around $U \approx 1.5$ and $U \approx 15$ which stretch up to
$\lambda \approx 2.4$.

The observed correlated metallic bound states have to form bands
with a continuous spectrum. The width of these bands will
determine the largest group velocity of spreading fronts as in
Fig.\ref{fig_2}. We found that these velocities are in general
depending on the control parameter values. In the following we
will discuss a possible mechanism for the observed effect of
appearance of correlated metallic bound states in the main
existence region of the correlated metal phase in Fig.\ref{fig_4a}
for $\lambda=2.5$. Let us consider the Fock space of two
noninteracting particle eigenfunctions
\begin{equation}
|\mu,\nu\rangle=\frac{1}{\sqrt{\delta_{\mu\nu}+1}}\sum\limits_{k,\ell}A_k^\mu
A_\ell^\nu\hat b_k^+\hat b_\ell^+|0\rangle.\label{eq6}
\end{equation}
where $\mu \leq\nu$ are the indices of the single particle eigenstates, sorted according
to their position along the chain. For nonzero interactions
we expand an eigenstate as
\begin{equation}
|\Psi(t)\rangle=\sum\limits_{\mu\leq \nu} \sum\limits_\nu
\phi_{\mu\nu}(t)|\mu,\nu\rangle \label{eq14}
\end{equation}
and the evolution equation for the coefficients ${\phi}_{\mu\nu}$
reads
\begin{equation}
i\dot{\phi}_{\mu\nu}={\cal
E}_{\mu\nu}{\phi}_{\mu\nu}+\sum\limits_{\mu'\leq\nu',\nu'} U\cdot
I_{\mu\nu,\mu'\nu'}{\phi}_{\mu'\nu'} \;. \label{eq1511}
\end{equation}
The renormalized energies ${\cal E}_{\mu\nu}$
and the overlap integrals $I_{\mu\nu,\mu'\nu'}$
responsible for hopping between the modes $|\mu,\nu\rangle$ and
$|\mu',\nu'\rangle$, are obtained from
\begin{eqnarray}
&{\cal E}_{\mu\nu}=(\lambda_\mu+\lambda_\nu)+U I_{\mu\nu}^0, \quad
I_{\mu\nu}^0=\frac{2}{\delta_{\mu\nu}+1}\sum_{j}\left(A_j^\mu
A_j^\nu\right)^2 \nonumber \\
&I_{\mu\nu,\mu'\nu'}=\frac{2}{\sqrt{(\delta_{\mu\nu}+1)(\delta_{\mu'\nu'}+1)}}
\sum_{j}A_j^\mu A_j^\nu A_j^{\mu'}A_j^{\nu'} \;. \label{eq15}
\end{eqnarray}

We are interested only in bound states where the two particles are within a localization length
distance from each other, since these are observed to yield a transition to a correlated metal.
The renormalization in each miniband depends on the strength of the overlap integrals $I^0_{\mu\nu}$.
We compute them numerically and find that the average overlap integrals $I^0\approx 0.5$ from TP1
and $I^0 \approx 0.3$ from TP3.
Note that the values for $I^0$ for states from TP2 and TP4 are much smaller, since the Fock states
are made of products of {\sl different} single particle states. TP5 yields again large values of $I^0$
but is irrelevant for reasons given below.
We also obtain that the average overlap integrals
$ I_{\mu\nu,\mu'\nu'} \approx 0.1$.
Let us consider only states from TP1. The increase in $U$
leads to a {\sl broadening} of TP1 width, since some energies get strongly renormalized and some less.
In the Fock space we therefore observe an increase of an effective potential strength (which is
similar to
$\lambda$ for the noninteracting case) as $0.5 U$. At the same time the different Fock states from
TP1 increase their overlap (which is similar to the hopping strength of a single particle) as
$0.1 U$. Therefore the increase of the potential strength wins, and these states do not cross over
into a delocalized regime, when no further Fock states are considered. The same is essentially true
for all TIP minibands. However, at some value of $U$ some renormalized states from TP1 will resonate with
weakly renormalized states from TP2. This group of states is characterized by a zero potential
strength (since they are resonant) and any finite overlap will therefore lead to a complete delocalization.
The expected value of $U$ follows from the distance between the minibands which is
around 2, and with $I^0 =0.5$ we predict the delocalization to start around $U=4$ as observed in
the numerics. This is the rough location of the left large tongue in Fig.\ref{fig_4a}.

Using the same reasoning we predict that a resonant mixing of renormalized states from TP3 with ones from TP4
is expected at around $U=6.7$, as follows from the miniband distance around 2 and the TP3 value $I^0 = 0.3$.
Again this is indeed the observed location of the right large tongue in Fig.\ref{fig_4a}.
It is a challenging task to extend the above arguments to the whole phase diagram in Fig.\ref{fig_4a}.

The novel state of a correlated metal formed from two interacting particles should be easily measured
using interacting pairs of ultracold Rb atoms in optical lattices, which were shown to bind despite their repulsive
interaction \cite{Zoller}.

To summarize, we observed a nonperturbative delocalization of two
interacting particles in a quasiperiodic potential deep in the
insulator phase of the noninteracting problem. The corresponding
correlated metallic bound states keep both interacting particles
at distance less than the single particle localization length.
This happens because the interacting particles may redistribute
their total energy into different Fock states which are coupled
due to the interaction. We gave estimates for the appearance of
the correlated metallic phase in the parameter space, which agree
well with the numerical data. It is a challenging task to extend
the theory to the whole parameter space, to make it quantitative,
and to explore the effect of further increase of particle number.
At finite particle density we expect to observe a many body
correlated metallic phase.
\\
\\
Acknowledgements
\\
We thank B. L. Altshuler and I. L. Aleiner for stimulating
discussions. MI acknowledges financial support of the Dynasty
Foundation, Russian Federation Government (contracts
11.G34.31.0066 and 14.740.11.0075) and RFBR 10-02-00865.

\end{document}